\UseRawInputEncoding
\documentclass{vgtc}                          % final (conference style)
\ifpdf%                                % if we use pdflatex
  \pdfoutput=1\relax                   % create PDFs from pdfLaTeX
  \usepackage{graphicx}                % allow us to embed graphics files
  \DeclareGraphicsExtensions{.pdf,.png,.jpg,.jpeg} % for pdflatex we expect .pdf, .png, or .jpg files
\else%                                 % else we use pure latex
  \usepackage{graphicx}                % allow us to embed graphics files
  \DeclareGraphicsExtensions{.eps}     % for pure latex we expect eps files
\fi%

\graphicspath{{figures/}{pictures/}{images/}{./}} % where to search for the images

\usepackage{microtype}                 % use micro-typography (slightly more compact, better to read)
\PassOptionsToPackage{warn}{textcomp}  % to address font issues with \textrightarrow
\usepackage{textcomp}                  % use better special symbols
\usepackage{mathptmx}                  % use matching math font
\usepackage{times}                     % we use Times as the main font
\usepackage{cite}                      % needed to automatically sort the references
\usepackage{tabu}                      % only used for the table example
\usepackage{booktabs}                  % only used for the table example
\usepackage{tabularray}
\usepackage{graphicx}
\usepackage{orcidlink}

\onlineid{0}

\vgtccategory{Research}

\vgtcinsertpkg

\title{Who's Watching Me?: Exploring the Impact of Audience Familiarity on Player Performance, Experience, and Exertion in Virtual Reality Exergames}

\author{Zixuan Guo \orcidlink{0000-0002-0451-8988}\\ %
       \parbox{1.4in}{\scriptsize \centering Xi'an Jiaotong-Liverpool University} %
\and Wenge Xu \orcidlink{0000-0001-7227-7437}\\ %
    \parbox{1.4in}{\scriptsize \centering Birmingham City University} %
\and Jialin Zhang \orcidlink{0000-0001-6044-8585}\\ %
    \parbox{1.4in}{\scriptsize \centering Xi'an Jiaotong-Liverpool University} %
\and Hongyu Wang \orcidlink{0000-0002-2288-5116}\\ %
    \parbox{1.4in}{\scriptsize \centering Xi'an Jiaotong-Liverpool University} %
\and Cheng-Hung Lo \orcidlink{0000-0002-7199-9339}\\ %
    \parbox{1.4in}{\scriptsize \centering Xi'an Jiaotong-Liverpool University} %
\and Hai-Ning Liang~\orcidlink{0000-0003-3600-8955}\thanks{Corresponding author (e-mail: haining.liang@xjtlu.edu.cn)}\\ %
    \parbox{1.4in}{\scriptsize \centering Xi'an Jiaotong-Liverpool University}}
     
\abstract{Familiarity with audiences plays a significant role in shaping individual performance and experience across various activities in everyday life. This study delves into the impact of familiarity with non-playable character (NPC) audiences on player performance and experience in virtual reality (VR) exergames. By manipulating of NPC appearance (face and body shape) and voice familiarity, we explored their effect on game performance, experience, and exertion. The findings reveal that familiar NPC audiences have a positive impact on performance, creating a more enjoyable gaming experience, and leading players to perceive less exertion. Moreover, individuals with higher levels of self-consciousness exhibit heightened sensitivity to the familiarity with NPC audiences. Our results shed light on the role of familiar NPC audiences in enhancing player experiences and provide insights for designing more engaging and personalized VR exergame environments.%
} % end of abstract

\CCScatlist{
  \CCScatTwelve{Human-centered computing}{Human computer interaction (HCI)}{Interaction paradigms}{Virtual reality};
  \CCScatTwelve{Software and its engineering}{Software organization and properties}{Virtual worlds software}{Interactive games}
}

\begin{document}


\firstsection{Introduction}

\maketitle

%% \section{Introduction} %for journal use above \firstsection{..} instead
Virtual Reality (VR) exergames have gained significant popularity as an innovative approach that combines physical exercise with immersive gaming experiences \cite{xu2020results,xu2021effect}. As sedentary lifestyles have become increasingly prevalent, with prolonged sitting leading to various health concerns, VR exergames offer a promising solution by promoting physical movement and mitigating the negative effects associated with excessive sitting \cite{zuo2023features}. These interactive virtual environments allow users to engage in physical activities while being transported to fun and captivating virtual worlds \cite{xu2020hiit,xu2022acceptance}. 

Non-Player Characters (NPC) and their feedback have been used in games to enhance gameplay experiences \cite{haller2019hiit, yu2023cheer}. However, there is little research on the mechanisms for establishing meaningful connections between these audiences and players, particularly in VR exergames. Tice et al. \cite{tice1995modesty} argued that how individuals present themselves is influenced by their audience; specifically, individuals may behave differently when facing familiar and unfamiliar audiences \cite{weary1981attributional, nesbit2019role}. Furthermore, several studies \cite{reis2011familiarity, burkitt2019expressivity} have supported the positive influence of familiar audiences on individual performance, as encountering familiar individuals has been shown to evoke a sense of connection, comfort, and heightened engagement. Conversely, encountering unfamiliar individuals may elicit heightened nervousness or anxiety, but sometimes it can also stimulate individuals to have a stronger desire to perform and excel \cite{leary1994self}. Research suggests that the extent of an audience's influence on users depends on an individual's level of self-consciousness \cite{baumeister1984choking, carver1978self, wang2004self}. Individuals with high self-consciousness are more concerned about others' judgments and experience increased anxiety when being observed, which can affect their performance and emotions.

A familiar audience has been shown to enhance social presence and increase user motivation and enjoyment in various contexts, including virtual environments. For instance, recent studies \cite{monteiro2023effects, barrett2023exploring} have demonstrated that when delivering a speech in a VR environment and facing familiar audiences, users tend to experience a greater sense of relaxation and exhibit improved performance. However, the impact of audience familiarity on player experiences in gaming contexts remains relatively unexplored. 

Moreover, the role of voice familiarity has been largely overlooked. Notably, audience familiarity can be perceived through different patterns of signals originating not only from facial features but also from voices \cite{young2020face}. Familiar appearances and voices can convey identity information and are believed to create impressions of warmth, competence, and other social characteristics \cite{ellis1997intra, schweinberger2014speaker}. Understanding the influence of audience familiarity in VR exergames is essential for creating engaging and personalized experiences. Therefore, the primary objective of this work is to explore how familiarity with NPC audiences, encompassing both appearance (facial features and body shape) and voice, impacts players' performance, experience, and exertion within the context of VR exergames. Specifically, we seek to answer the following research questions:

\begin{itemize} 
\item [\textbf{RQ1:}] How would the familiar/unfamiliar appearances of NPC audiences affect players' performance, experience, and exertion in VR exergames?
\item [\textbf{RQ2:}] How would the familiar/unfamiliar voices of NPC audiences impact players' performance, experience, and exertion in VR exergames?
\item [\textbf{RQ3:}] How does the level of self-consciousness relate to the impact of audience familiarity on players' performance, experience, and exertion in VR exergames?
\end{itemize}

To answer these RQs, we conducted a 2 $\times$ 2 within-subjects experiment (1. Audience Appearance---unfamiliar and familiar and 2. Audience Voice---unfamiliar and familiar) with 24 young adults participating in a VR exergame. Our findings point to the positive impact of both familiar audience appearance and voice on players' overall gaming experience, emotional state, and perceived exertion. Furthermore, the familiar voice enhances player performance in the game. Additionally, individuals with higher levels of self-consciousness demonstrate a heightened sensitivity to the familiarity with NPC audiences, leading to a greater sense of relaxation when facing familiar audiences. We provide two design recommendations incorporating audience familiarity to support designers and developers in shaping players' game performance and experience, ultimately contributing to a healthier lifestyle.

\section{Related work}
%\subsection{VR Exergames}
%VR exergames are becoming increasingly well-liked in the gaming industry and academic community. For example, commercial games like Virtual Sports3 enables players move their entire body or use gestures to play various sports. In FitXR4, the players need to jab, weave, and do uppercuts following  the music rhythm. Previous research on VR exergames has focused on improving the player's experience and exertion by improving interactions, including display modes \cite{yoo2016vrun, xu2020studying}, design of bodily gestures \cite{ioannou2019virtual, xu2020results}, gameplay \cite{xu2021effect, liu2020virtual}, etc. For instance, Yoo and Kay \cite{yoo2016vrun} demonstrated VRun to encourage players to run, and found that the VR environment can provide players with more immersion and motivation to exercise than large wall display and baseline laptop. Ioannou et al. \cite{ioannou2019virtual} designed and implemented a VR exergame that can virtually enhance the player's performance, allowing players to feel an illusion of greater ability than they actually have during running and jumping. Xu et al. \cite{xu2021effect} introduced uncertain factors into a fighting-based VR exergame, and practically proved that uncertain factors can help improve exertion levels.

%Yet, VR exergames have other elements that may also be crucial for a positive gaming experience and enhanced performance. According to real physical activities, the presence of audiences may be one of the important reasons affecting the performance of participants.

\subsection{NPC Audience and Its Impact on Users}

%Bowman et al. \cite{bowman2013facilitating} suggested that the presence of a physical audience during gameplay was a significant predictor of enhanced game performance. 
%Kappen et al. \cite{kappen2014engaged} emphasized the attitude of audiences, and the results showed that both passive and active audiences increase player engagement, but silent audiences make players feel unnerved.
NPC audiences can have a great impact on users' performance and experience. For instance, Pertaub et al. \cite{pertaub2002experiment} found that negative virtual audiences triggered heightened anxiety levels among users during public speaking, in contrast to positive and neutral audiences. Yakura and Goto \cite{yakura2020enhancing} discovered that when users are part of the concert audience, the behavior of other concert NPC audiencs could significantly enhance their sense of co-presence and improve their overall concert experience. Strojny et al. \cite{audienceVRFire} studied VR training for firefighter first response and found that social presence plays a crucial role in moderating the impact of virtual audiences on performance in such scenarios. Kabeko et al. \cite{kaneko2018supporting} found NPCs' feedback can enhance the user's experience and create a sense of ``unity" with the audiences in a VR live house. Raimbaud et al. \cite{raimbaud2022stare} investigated the gaze behavior of virtual audiences in VR, revealing that participants demonstrated quicker detection and prolonged gaze duration in response to directed gaze, mirroring real-world tendencies.

Prior research has also explored the influence of NPC audiences on video games. For example, Haller et al. \cite{haller2019hiit} suggested that feedback from a virtual audience could lead to better performance and higher heart rates in players. Xu et al. \cite{xu2023exploring} found that the group size of an NPC audience and its feedback could positively affect a player's performance and experience while playing VR exergames. Focusing on older players, Yu et al. \cite{yu2023cheer} found that virtual audiences with reactive feedback could enhance their performance and experience in VR exergames. Yet, no studies have focused on the investigation of audience familiarity specifically in the context of VR games, including VR exergame. This work fills this gap.

\subsection{Users' Sense of Familiarity and Its Impact}
From a psychological standpoint, familiarity can be defined as the level of knowledge or acquaintance that an individual possesses regarding specific stimuli, such as physical stimuli, environmental elements, and other individuals \cite{goodman1991familiarity}. For other individuals, familiarity can be perceived on a continuum, ranging from complete strangers, and acquaintances, to close friends and family. 

Existing research has demonstrated a link between sensory familiarity and the experience of positive affect \cite{harmon2001role, garcia2010familiar, garcia2016familiarity}. For visual stimuli, Harmon-Jones and Allen \cite{harmon2001role} found that participants exhibited greater activity in facial muscles associated with positive affect after being exposed to familiar photos than unfamiliar photos. Garcia-Marques et al. \cite{garcia2010familiar} demonstrated that familiarity and positive affect influence each other's judgment latencies through a symbol judgment task. In the realm of auditory stimuli, Garcia-Marques et al. \cite{garcia2016familiarity} delved into the evaluation of recorded statements and unveiled that the processing of familiar stimuli resulted in a pleasurable subjective experience.

When confronted with stimuli beyond simple sensory input, such as familiar people, individuals can also demonstrate more positive performance and psychological states. For example, MacIntyre and Thivierge \cite{macintyre1995effects} found that in public speaking, facing familiar audiences can make participants perceive stronger agreeableness and willingness to speak and lower anticipated anxiety than unfamiliar audiences. Burkitt et al. \cite{burkitt2019expressivity} investigated the effect of audience familiarity on children's expressive drawings of themselves and found that children showed more positive expressiveness when faced with familiar peers and adult audiences. Monteiro et al. \cite{monteiro2023effects} investigated the influence of audience familiarity in a VR public speaking training tool and found that participants with medium public speaking anxiety exhibited a significantly more relaxed demeanor when faced with familiar NPC audiences. 

These studies provide evidence that familiar audiences have the potential to induce a more relaxed mood and facilitate participants' performance. However, there is currently limited research that incorporates the familiarity with the NPC audiences within virtual gaming environments. Furthermore, our understanding of the influence of visual and auditory stimuli presented by audiences remains incomplete. To address these gaps, this study aims to investigate how familiarity with the NPC audience's appearance (including face and body shape) and voice influence players' performance and overall experience in VR exergames.

\subsection{Self-consciousness}
While the presence of an audience is often believed to enhance individual performance, it is essential to consider the role of self-consciousness, as it can significantly impact how individuals respond to audience interactions \cite{belletier2019social}. Self-consciousness refers to an individual's awareness and attention to their own thoughts, feelings, and behaviors in relation to others \cite{fenigstein1975public}. It involves a focus on how one is perceived by others and a concern for social evaluation. 

Several studies \cite{baumeister1984choking, carver1978self, wang2004self} have explored the relationship between self-consciousness and the influence of audiences on personal performance. For example, Wang et al. \cite{wang2004self} suggested that individuals with high levels of self-consciousness may experience heightened anxiety and decreased performance in front of an audience. This effect appears to be more pronounced when individuals are confronting unfamiliar audiences. Mesagno et al. \cite{mesagno2012choking} observed that basketball players with high self-consciousness showed significantly increased anxiety and diminished performance when facing unfamiliar audiences. In these instances, the presence of an audience may lead to self-doubt and increased self-monitoring, which can disrupt performance and hinder task execution \cite{baumeister1984choking}. Existing research \cite{kappen2014engaged, haller2019hiit, yu2023cheer} on audience familiarity in virtual environments has paid limited attention to users' self-consciousness. Considering that self-consciousness may be a crucial influencing factor, we aim to explore the relationship between players' self-consciousness and their exposure to NPC audience familiarity in this work.

% Moreover, self-consciousness also responds differently to familiar and unfamiliar situations. Wilhelm et al. \cite{wilhelm2001social} found that individuals with high self-consciousness reported higher levels of anxiety and embarrassment in social situations involving unfamiliar individuals. When facing unfamiliar audiences, Mesagno et al. \cite{mesagno2012choking} observed that basketball players with high self-consciousness showed significantly increased anxiety and decreased performance when facing unfamiliar audiences.

\section{Exergame Application}

We have developed a VR exergame, which is inspired by Fruit Ninja, using the Unity3D engine. Expanding on the base game, our design includes a fog interference feature, encouraging players to squat and clear the fog, boosting exertion, and adding a challenging aspect. Furthermore, our game includes virtual audiences to enrich the player experience. All variables reported in this section were decided through several playtesting sessions with four game testers. 

\begin{figure}[htbp]
    \centering
    \includegraphics[width=\columnwidth]{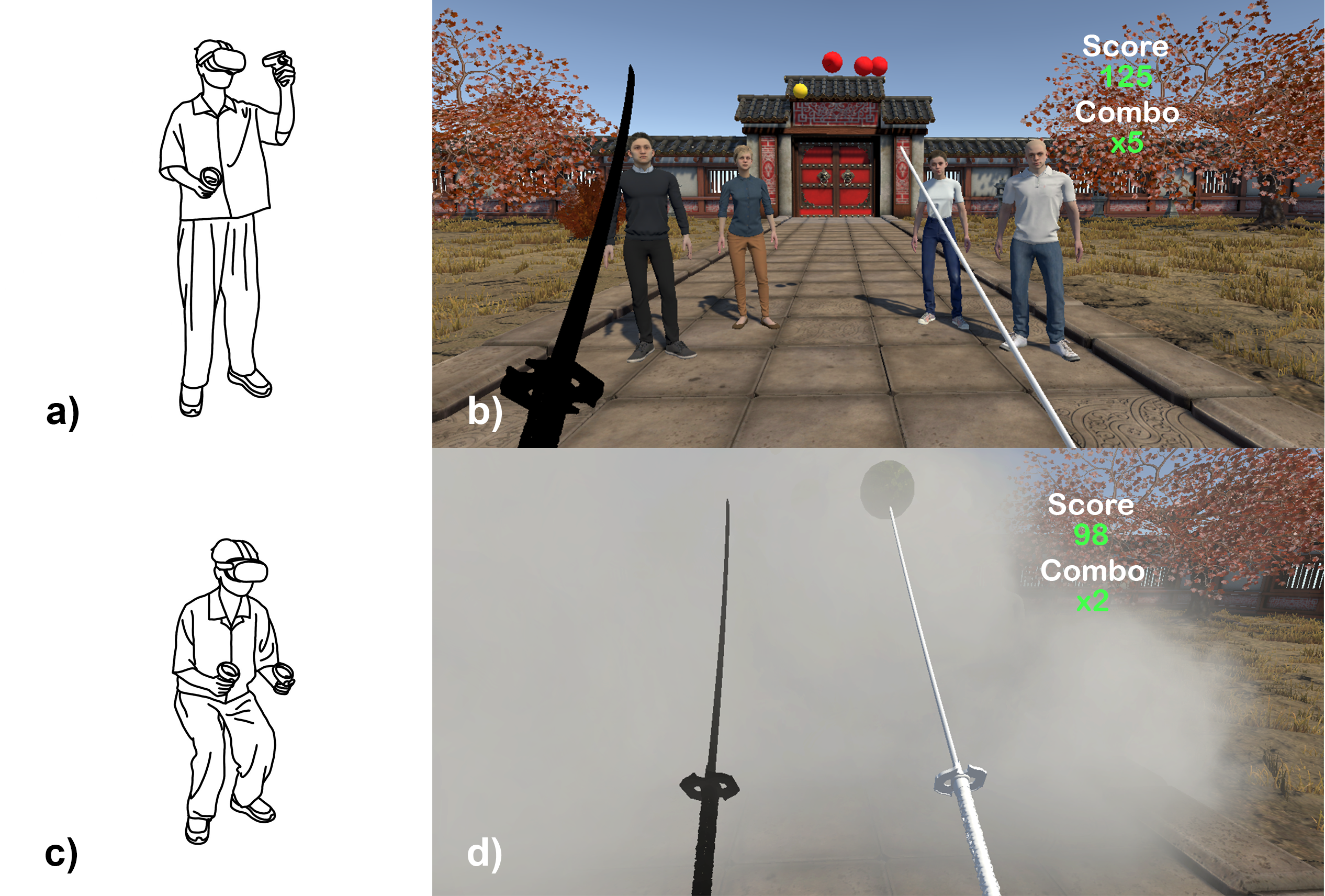}
    \caption{(a) Real and (b) Virtual perspectives of fruit slicing gameplay using controllers. (c) Real and (d) Virtual perspectives of squatting to clear fog interference.}
    \label{gameplay}
    
\end{figure}

\subsection{Rules and Logic}

During the game, players are required to use two hand-held controllers to wield virtual swords to slice as many fruits (watermelon, apple, and lemon) as possible and dodge bombs. 
Figure \ref{gameplay} shows that fruits and bombs are launched towards the left and right sides of the player, following a parabolic trajectory that enables them to land within the reachable area of the player's virtual sword. To ensure trajectory reliability, we first adjusted the gravity setting based on feedback from the four game testers of different heights (160cm - 185cm), thereby extending the flight time of the targets. We also adjusted the initial velocities on the X, Y, and Z axes to restrict the angle range, allowing participants to hit the targets within a consistent area of approximately one square meter.

The game lasts 5 minutes and 12 seconds, composed of 5 one-minute game sequences with 3 seconds of rest time between each game sequence. Each one-minute sequence consists of 20 3-second rounds that generate a fixed number of fruits and bombs in random order. Each round contains 2-4 fruits, with a bomb appearing every 4-5 rounds. In total, there are 100 rounds (20 per minute), and the occurrences of each target are as follows: (1) watermelon: 90, (2) apple: 80, (3) lemon: 85, and (4) bomb: 20. 

The game also includes a fog interference feature which obstructs the player's view to elevate the challenge and variation of the gameplay. The player needs to squat to disperse the fog and regain their vision (Figure \ref{gameplay}). We utilized the Steam effects from Unity's Particle pack to create the fog, with a simulation speed of 1 and a resulting shape approximating a sphere with a diameter of 1.8 meters. We placed four white-gray smoke effects around the camera to completely obstruct the player's view. The fog generation frequency was set at 6-10 seconds, resulting in the appearance of fog approximately 25-35 times per game, based on the player's 3-second reaction and squatting duration. This ensured the fog interference mechanism was sufficiently challenging without compromising the players' overall gaming experience.

\subsection{Audience Setting}

\begin{figure*}[htbp]
    \centering
    \includegraphics[width=\textwidth]{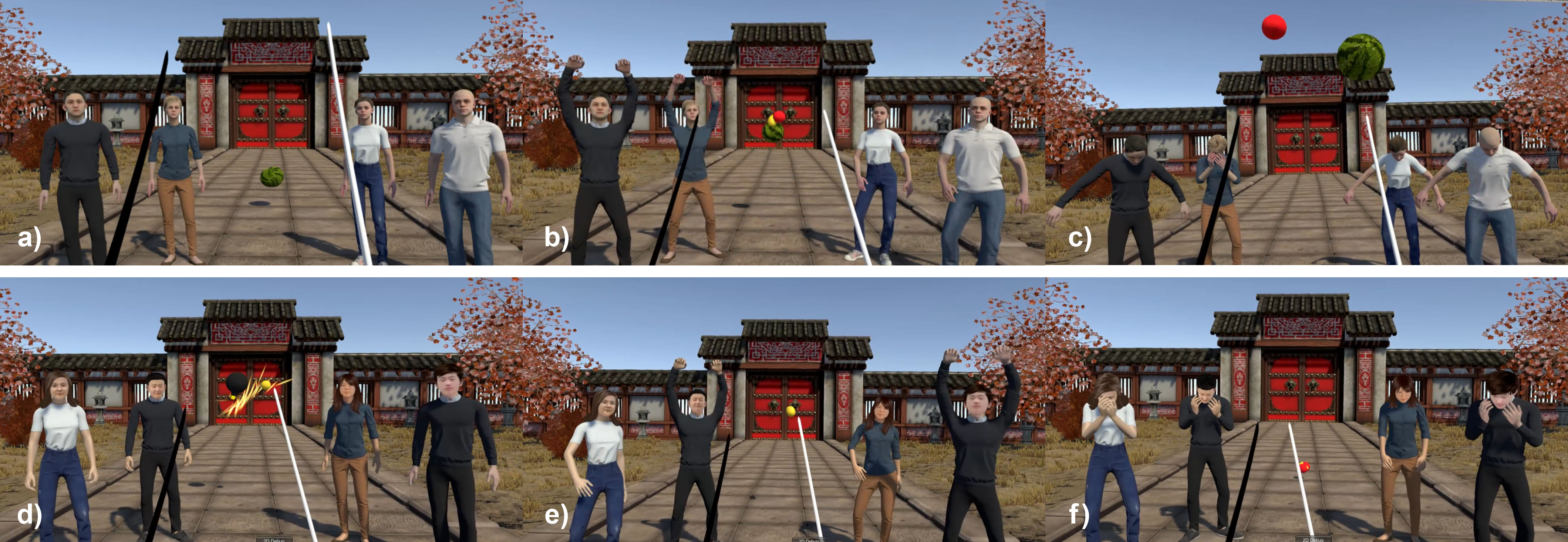}
    \caption{Top Row: (a) Unfamiliar NPC audiences, (b) Unfamiliar NPC audiences providing positive feedback, (c) Unfamiliar NPC audiences expressing negative feedback.
Bottom Row: (d) Familiar NPC audiences, (e) Familiar NPC audiences providing positive feedback, (f) Familiar NPC audiences expressing negative feedback.}
    \label{audience}
    
\end{figure*}

To ensure optimal visibility of each NPC audience's faces, we positioned four NPCs as audiences in the player's field of view. These NPCs provided both positive and negative feedback to the player, with each feedback consisting of two different animations and two voice lines (Figure \ref{audience}). The two positive feedback animations include raising both hands and jumping while waving one hand upward when cheering for the player. The two negative feedback animations consist of covering the face with both hands and lowering the head while waving one hand downward. The positive voice lines are delivered with excitement and joy, while the negative voice lines convey a sense of disappointment and frustration. To enable participants to identify the voices of the NPCs, their voice feedback was set to first emit some exclamations (such as ``Oh" and ``Wow") together, and then a random audience member played a voice line with content (such as ``Great job" and ``What a pity"). The four NPCs provided feedback simultaneously, but the specific animations and voice lines played are randomized to prevent predictability. 

\subsection{Scoring System And Feedback Activation}
As shown in Table \ref{Scoring Table}, players received three types of scores: total score, combo count, and rating score. The total score shows on the upper right corner of the gameplay screen and reflects the total points gained at the end of the gameplay. Combo count tracked the consecutive number of sliced fruits, which is presented below the total score display. The rating score with rating level served as a trigger for NPC feedback and was not visible to players. The rating score and rating level both start at 0. Each level is separated by a 10-point interval, which is determined by dividing the current rating score by 10 and taking the integer part. When players reach a higher level, it triggers positive feedback, while descending to a lower level triggers negative feedback. There is no upper limit to the rating level, but the lower limit is set at 0. If a player remains at level 0 for over 6 seconds, it also triggers negative feedback. To prevent excessive disruptions, a 4.5-second cooldown period followed each feedback event, during which no additional feedback was triggered. 

\begin{table}
\caption{Scoring Metrics for Fruits and Bombs - Total Score, Combo Count, and Rating Score}
\centering
\scalebox{0.85}{
\begin{tblr}{
  cells = {c},
  cell{2}{1} = {r=2}{},
  cell{4}{1} = {r=2}{},
  cell{6}{1} = {r=2}{},
  hline{1-2,4,6,8-9} = {-}{},
}
Item       & State  & Total Score & Combo Count & Rating Score \\
Watermelon & Sliced & +3          & +1          & +5           \\
           & Missed & -7          & Reset to 0  & -5           \\
Apple      & Sliced & +5          & +1          & +5           \\
           & Missed & -5          & Reset to 0  & -5           \\
Lemon      & Sliced & +7          & +1          & +5           \\
           & Missed & -3          & Reset to 0  & -5           \\
Bomb       & Hit    & -10         & Reset to 0  & -10          
\end{tblr}
}
\label{Scoring Table}
\end{table}

\section{Experiment}
\subsection{Experiment Design}
The study employed a 2 $\times$ 2 within-subjects design that incorporated two factors: Audience Appearance (unfamiliar and familiar) and Audience Voice (unfamiliar and familiar). In total, we have evaluated 4 conditions: (1) Unfamiliar Appearance with Unfamiliar Voice (UAUV), (2) Unfamiliar Appearance with Familiar Voice (UAFV), (3) Familiar Appearance with Unfamiliar Voice (FAUV), and (4) Familiar Appearance with Familiar Voice (FAFV). The order of these conditions was counterbalanced in the experiment. 

When creating familiar audiences, the familiar face was obtained via scanned photographs of participants' friends (using the Avatar Maker Pro plugin), and the body shape and height were generated by adjusting the size of the models. The same individuals also provided familiar voices through voice recordings. On the other hand, the unfamiliar audiences were created using realistic computer character models, and the unfamiliar voices were generated by Microsoft Azure, an AI voice generator that produces natural and smooth human speech based on text. Other factors such as the phrases used for the voices, physical characteristics of the NPC audiences (adjusted to suit the individuals known to the player), and the number of NPCs were controlled in all conditions.

\subsection{Outcome Measures}

\begin{itemize}

\item \emph{\textbf{Player Performance.}} We collected the following performance measurements: (1) game score; (2) success rate of fruit hit and avoiding bomb; (3) count of positive and negative feedback; and (4) maximum combo count. See Section 3 for a detailed description. 

\item \emph{\textbf{Player Experience.}} We employed the following three scales: 
\begin{itemize}
    \item Player Experience of Need Satisfaction (PENS) scale \cite{ryan2006motivational}: a 21-item scale designed to measure game experience across five dimensions. (1) Competence: the players' perception of their own capabilities. (2) Autonomy: the extent to which players felt free to make choices within the game. (3) Relatedness: the players' sense of connection to "other players in the game," (i.e., the NPC audience in our study). (4) Presence/Immersion: players' emotional engagement in the game. (5) Intuitive Controls: the degree to which players feel that their choices translate smoothly and seamlessly into in-game actions. 
    \item Pick-A-Mood (PAM) tool \cite{desmet2016mood}: a cartoon-based pictorial instrument for quick and intuitive reporting and expression of mood states. This tool categorized eight distinct mood states into four main categories: Energized-Pleasant (excited and cheerful), Calm-Pleasant (relaxed and calm), Energized-Unpleasant (irritated and tense), and Calm-Unpleasant (bored and sad). 
    \item Simulator Sickness Questionnaire (SSQ) \cite{kennedy1993simulator}: a commonly used 16-item self-report questionnaire that produces 3 measures of cybersickness (nausea, oculomotor, and disorientation). A total SSQ score ranging from 20 to 30 indicates mild to moderate simulator sickness, while a score exceeding 40 is indicative of a poor simulator experience \cite{caserman2021cybersickness}. 
\end{itemize}

%The Pick-A-Mood (PAM) tool \cite{desmet2016mood} consists of four categories: Energized-Pleasant, Calm-Pleasant, Energized-Unpleasant, and Calm-Unpleasant. 

\item \emph{\textbf{Exertion.}} The exertion was assessed by (1) the average heart rate (AvgHR\%) using the formula the average heart rate / the participant's estimated maximum heart rate (i.e., 211-0.64×age \cite{nes2013age}); (2) calories burned; and (3) the Borg Rating of Perceived Exertion (RPE) 6-20 scale \cite{borg1982psychophysical}.

\item \emph{\textbf{Self-Consciousness Scale (SCS).}} The SCS \cite{fenigstein1975public} consisted of three subscales (private self-consciousness, public self-consciousness, and social anxiety), and the scores from the three subscales are summed to obtain an individual's total self-consciousness score. Private self-consciousness refers to an individual's tendency to introspect and reflect on their inner thoughts, feelings, and experiences. Public self-consciousness, on the other hand, refers to an individual's awareness of how they are perceived by others in social situations. Social anxiety is the tendency to experience discomfort and anxiety in social situations.

\item \emph{\textbf{Ranking and Interviews.}} We asked participants to rank the evaluated conditions and provide reasons for their rankings. Then, we employed a semi-structured interview with the following open-ended questions: (1) ``How do you think the NPCs' familiarity affects your overall performance of the game?"; (2) ``How do you think the NPCs' familiarity affects your overall game experience?"; (3) ``Which factor of the NPCs do you think has a greater impact on your game experience, the appearance or the voice?"; (4) ``How do you think positive and negative feedback from the NPCs affect you?"; and (5) ``Would having familiar NPCs in future VR exergames encourage you to continue playing in the long term?". Interviews were recorded and transcribed for data analysis.

\end{itemize}

\subsection{Participants}
We recruited 24 participants (8 females; mean age = 23.00, SD = 2.02, range = 20 to 28) through an on-campus social media platform. Out of the 18 participants who reported having prior experience with VR HMDs, 9 were frequent weekly users. 11 of them had played exergames before, but none were regular players. Regarding how engaged they were with daily exercise, 5 participants exercised regularly (more than 3 hours per week), 13 participants were less regular (1-3 hours per week), and 6 were inactive (less than 1 hour per week).

\subsection{Apparatus and Setup}
The experiment used a Meta/Oculus Quest 2 as the VR device and a Polar OH1 for monitoring participants' HR and calories burned. The experiment was conducted in a controlled laboratory environment that was well-illuminated and isolated from external disturbances. The room temperature was regulated by an air conditioner that maintained a constant temperature of 24℃ during the experiment. The experiment was conducted under close supervision by an experimenter to prevent any potential risks. This experiment received approval from the University Ethics Committee at the host institution.

%, which is linked to a computer with an i7 CPU, 32 GB RAM, and NVIDIA RTX 2080 Ti GPU using a Meta Quest Link cable

\begin{figure*}[htbp]
    \centering
    \includegraphics[width=\textwidth]{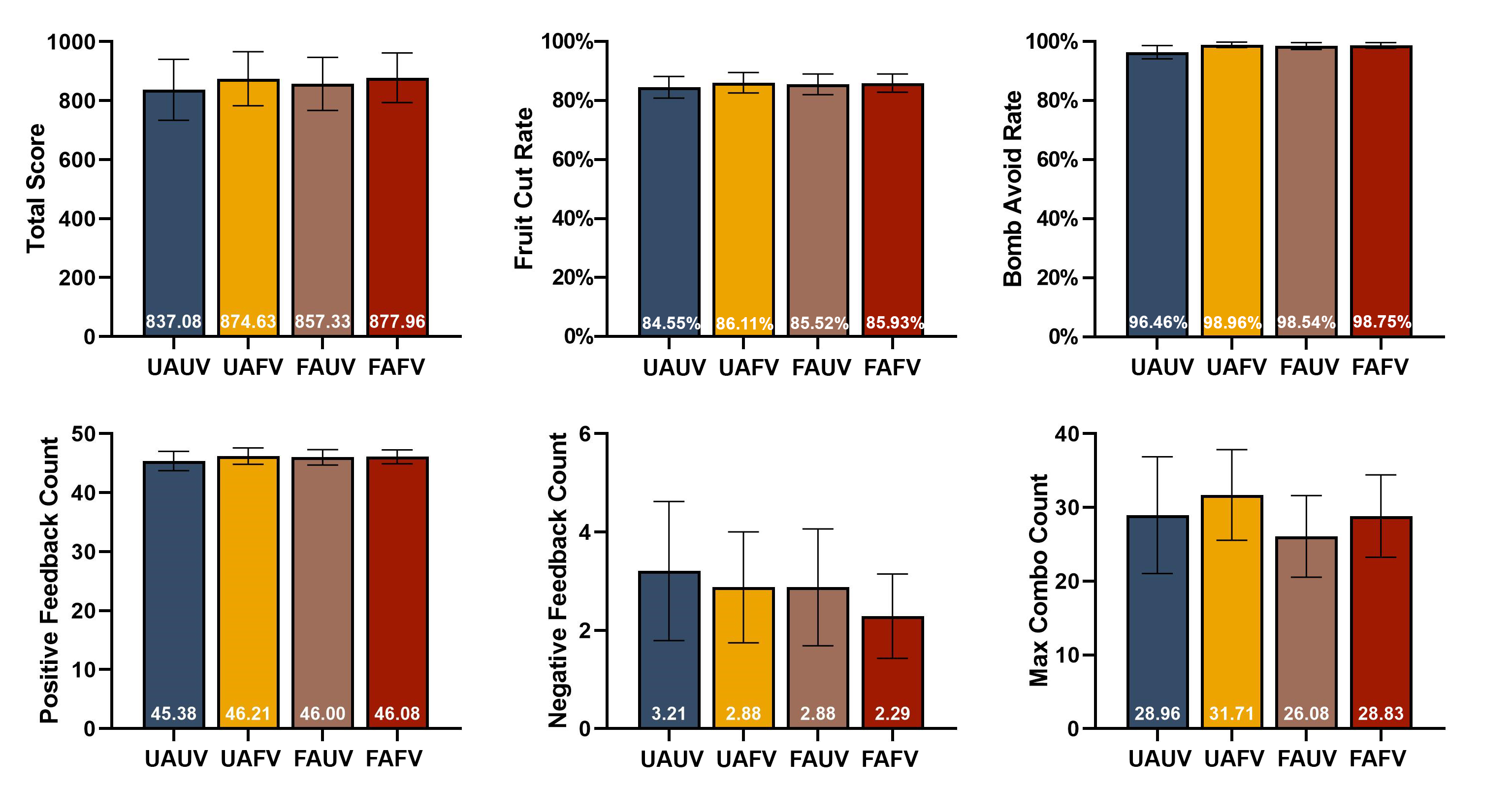}
    \caption{Mean total score, fruit cut rate, bomb avoidance rate, positive feedback count, negative feedback count, and maximum combo count for each condition. Error bars indicate 95\% confidence intervals.}
    \label{Performance}
    
\end{figure*}

\subsection{Procedures}
Participants were provided with a comprehensive introduction of the experimental protocol to read and a consent form to sign off prior to participating in the study. Then, they were required to bring 4 friends in a separate appearance and voice collection session to generate NPC before the experiment. Their friends would also need to sign the consent form to agree that they were willing to form the NPCs in the game. During the appearance and voice collection session, participants' friends were instructed (1) to maintain a neutral expression and to fully exhibit their facial features to generate the best facial expression, (2) to provide verbal feedback based on a list of prompts, which consisted of both exclamations (such as ``Oh" and ``Wow") and complete sentences (such as ``Great job", ``Well done", ``Oh no", and ``What a pity") \cite{epting2011cheers}. All photographs and voice recordings were collected using an iPhone 11. After creating the familiar NPCs, participants were asked to independently identify the models and voices of the NPCs to ensure their recognizability. No difficulties were reported during this process.

On the day of the experiment, participants needed to complete a pre-experiment questionnaire asking demographic questions including age, gender, frequency of exercise, experience with exergames and VR HMDs, as well as self-consciousness via SCS. Then, participants were required to enter their personal information (i.e., age, gender, height, and weight) into the Polar Beat mobile application. We collected the rest level HR using the Polar OH1 HR monitor by asking participants to relax and sit still for over 1 minute. 

Before the formal gameplay sessions, a 3-minute training phase without any NPCs was administered to enable participants to practice moves and become more acquainted with the equipment. For each experimental condition, participants were assisted by the experimenter to wear the Polar OH1 and HMD. The experiment commenced only when the participant's heart rate had reached the resting level. After each condition, participants needed to complete four questionnaires: PENS \cite{ryan2006motivational}, PAM \cite{desmet2016mood}, SSQ \cite{kennedy1993simulator}, and Borg RPE 6-20 \cite{borg1982psychophysical}. They then rested until they felt ready to begin the next experimental condition and had reached a resting heart rate level. At the end of the experiment, participants underwent a semi-structured interview to rank the experimental conditions and provide feedback on their experiences. The experiment lasted about 50 minutes for each participant.

\section{Results}
We conducted statistical analysis on the data using a two-way repeated measures ANOVA with Audience Appearance (unfamiliar and familiar) and Audience Voice (unfamiliar and familiar) as within-subjects variables. The normal distribution of the data was assessed using Shapiro-Wilks tests and Q-Q plots. For non-normally distributed data, we applied the Aligned Rank Transform (ART) to transform the data before performing repeated-measures ANOVAs. We utilized Bonferroni correction for pairwise comparisons, and we reported effect sizes whenever feasible ($\eta_p^2$). 

To analyze correlations between the participants' self-consciousness and their performance and experience, we conducted Pearson’s bivariate correlation analyses.

\begin{figure*}[htbp]
    \centering
    \includegraphics[width=\textwidth]{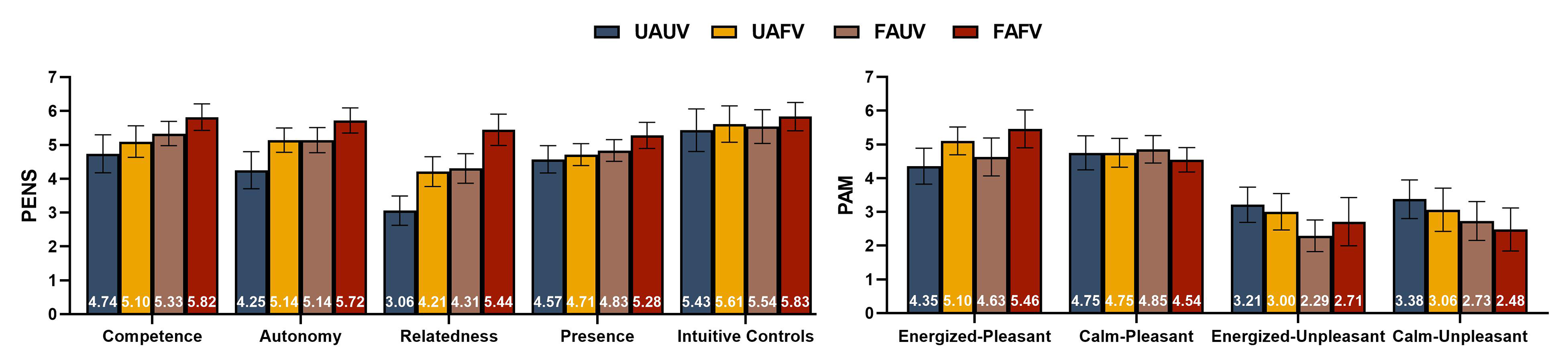}
    \caption{PEN and PAM ratings for each condition. Error bars indicate 95\% confidence intervals.}
    \label{PP}
    
\end{figure*}

\subsection{Player Performance}
Figure \ref{Performance} shows the performance measures for each condition. ANOVA tests yielded a significant main effect of Audience Voice on total score ($F_{1,23}=6.843, p=.015, \eta_p^2=0.229$), but not for Audience Appearance ($F_{1,23}=0.437, p=.515$) and their interaction effect ($F_{1,23}=0.209, p=.652$). Pairwise comparisons revealed that participants achieved higher scores when exposed to a familiar audience voice ($M=876.29, SE=40.50$) than unfamiliar audience voice ($M=847.21, SE=44.94$).

For fruit cut rate, there was a significant main effect of Audience Voice ($F_{1,23}=5.691, p=.026, \eta_p^2=0.198$), but no significant effects were observed for Audience Appearance ($F_{1,23}=0.346, p=.562$) and Audience Appearance $\times$ Audience Voice ($F_{1,23}=0.825, p=.373$). Subsequent pairwise comparisons revealed that participants exhibited a higher fruit cut rate when exposed to familiar audience voice ($M=86\%, SE=0.02$) in comparison to unfamiliar audience voice ($M=85\%, SE=0.02$).

The data of maximum combo count was not normally distributed and underwent an ART prior to conducting the RM-ANOVA. Our analysis suggested Audience Voice ($F_{1,23}=8.214, p=.009, \eta_p^2=0.263$) had a significant effect on player's maximum combo count, but not Audience Appearance ($F_{1,23}=1.849, p=.187$) and Audience Appearance $\times$ Audience Voice ($F_{1,23}=1.047, p=.317$). Post-hoc pairwise comparisons revealed that participants achieved a significantly higher maximum combo count in familiar voice conditions ($M=30.27, SE=2.63$) than in unfamiliar voice conditions ($M=27.52, SE=2.68$). No significant effects were found for bomb avoidance rate, positive feedback count, and negative feedback count.

Pearson's correlation unveiled a significant correlation between participants' self-consciousness and (1) the number of negative feedback they triggered when facing an unfamiliar audience appearance ($r=-.437, p=.033$); (2) their maximum combo count when confronted with a familiar audience appearance ($r=.414, p=.044$) and (3) familiar audience voice ($r=.422, p=.040$). No other significant correlations were found.

\subsection{Player Experience}
\subsubsection{Game Experience}

Values for each condition for PENS ratings can be found in Figure \ref{PP}. Concerning Competence, ANOVA analysis revealed significant main effects for Audience Appearance ($F_{1,23}=13.327, p=.001, \eta_p^2=0.367$) and Audience Voice ($F_{1,23}=14.266, p=.001, \eta_p^2=0.383$), while no significant interaction effect was found ($F_{1,23}=0.268, p=.610$). Participants reported a heightened sense of competence when (1) presented with a familiar audience appearance ($M=5.58, SE=0.16$) in contrast to an unfamiliar appearance ($M=4.92, SE=0.24$), (2) exposed to a familiar audience voice ($M=5.46, SE=0.18$) compared to unfamiliar voices ($M=5.04, SE=0.20$).

Our analysis found significant main effects of Audience Appearance ($F_{1,23}=22.255, p=.000, \eta_p^2=0.492$) and Audience Voice ($F_{1,23}=21.203, p=.000, \eta_p^2=0.480$) on Autonomy. No significant interaction effect of these two factors was found ($F_{1,23}=1.184, p=.288$). Participants expressed a stronger sense of autonomy (1) when facing a familiar audience appearance ($M=5.43, SE=0.15$) than an unfamiliar appearance ($M=4.69, SE=0.20$) and (2) when exposed to a familiar audience voice ($M=5.43, SE=0.15$) compared to unfamiliar voice ($M=4.69, SE=0.19$).

ANOVA tests indicated significant main effects of Audience Appearance ($F_{1,23}=50.013, p=.000, \eta_p^2=0.685$) and Audience Voice ($F_{1,23}=45.091, p=.000, \eta_p^2=0.662$) on Relatedness. No significant interaction effect was observed for Audience Appearance $\times$ Audience Voice ($F_{1,23}=0.003, p=.957$). Participants reported a stronger sense of relatedness (1) when encountering familiar audience appearance ($M=4.88, SE=0.18$) compared to unfamiliar appearance ($M=3.63, SE=0.19$) and (2) when exposed to familiar audience voice ($M=4.83, SE=0.19$) compared to unfamiliar voice ($M=3.68, SE=0.18$).

For Presence/Immersion, significant main effects were found for Audience Appearance ($F_{1,23}=16.152, p=.001, \eta_p^2=0.413$) and Audience Voice ($F_{1,23}=16.565, p=.000, \eta_p^2=0.419$), but not for Audience Appearance $\times$ Audience Voice ($F_{1,23}=3.871, p=.061$). Participants experienced a higher level of immersion (1) with a familiar audience appearance ($M=5.06, SE=0.16$) in contrast to an unfamiliar appearance ($M=4.64, SE=0.17$) and (2) when presented with a familiar audience voice ($M=5.00, SE=0.16$) compared to unfamiliar voice ($M=4.70, SE=0.17$). Notably, no main nor interaction effects were found for Intuitive Controls.

% \begin{figure}[htbp]
%     \centering
%     \includegraphics[width=8cm]{figures/Prism_Pens.png}
%     \caption{PENS}
%     \label{PENS}
    
% \end{figure}

Furthermore, Pearson's correlation revealed significant associations between participants' self-consciousness and their perceived Competence when faced with a familiar audience appearance ($r=.431, p=.036$) and familiar audience voice ($r=.424, p=.039$). Additionally, we found a significant correlation between participants' self-consciousness and their perceived Relatedness when exposed to familiar audience appearance ($r=.494, p=.014$) and familiar audience voice ($r=.460, p=.024$). No other significant correlations were observed.

\subsubsection{Emotion}
The ratings for PAM in each condition can be observed in Figure \ref{PP}. Regarding Energized-Pleasant, Our analysis showed a significant main effect of Audience Voice ($F_{1,23}=19.959, p=.000, \eta_p^2=0.465$), but not for Audience Appearance ($F_{1,23}=3.119, p=.091$) and interaction effects ($F_{1,23}=0.371, p=.548$). Participants reported a heightened sense of energized-pleasantness when exposed to a familiar audience voice ($M=5.28, SE=0.22$) as opposed to an unfamiliar audience voice ($M=4.49, SE=0.25$).

In terms of Energized-Unpleasant, ANOVA tests revealed a significant main effect of Audience Appearance ($F_{1,23}=9.543, p=.005, \eta_p^2=0.293$), but not for Audience Voice ($F_{1,23}=0.466, p=.502$) and interaction effects ($F_{1,23}=3.363, p=.080$). Participants expressed a stronger sense of energized-unpleasantness when confronted with an unfamiliar audience voice ($M=2.75, SE=0.21$) than a familiar audience voice ($M=2.85, SE=0.27$).

For Calm-Unpleasant, ANOVA tests yielded a significant main effect of Audience Appearance ($F_{1,23}=9.425, p=.005, \eta_p^2=0.291$), but not for Audience Voice ($F_{1,23}=3.955, p=.059$) and Audience Appearance $\times$ Audience voice ($F_{1,23}=0.070, p=.793$). Participants reported a greater degree of calm-unpleasantness when exposed to an unfamiliar audience voice ($M=3.05, SE=0.26$) compared to a familiar audience voice ($M=2.77, SE=0.28$).

No significant effects were observed for the Calm-Pleasant category. Moreover, our analysis did not reveal any significant correlations between the four emotional categories and participants' self-consciousness. %These findings indicate that participants' experiences of pleasantness and unpleasantness were not significantly influenced by their levels of self-consciousness.

% \begin{figure}[htbp]
%     \centering
%     \includegraphics[width=8cm]{figures/Prism_Pam.png}
%     \caption{PAM}
%     \label{PAM}
    
% \end{figure}

\subsubsection{Simulator Sickness}

None of our participants had an SSQ higher than 20 among four conditions---UAUV ($M=3.27, SE=3.87$), UAFV ($M=2.81, SE=4.01$), FAUV ($M=2.96, SE=4.41$), and FAFV ($M=2.96, SE=4.41$). Our analysis also did not show  significant results on Nausea, Oculomotor, Disorientation, as well as total SSQ scores.

\begin{figure*}[htbp]
    \centering
    \includegraphics[width=\textwidth]{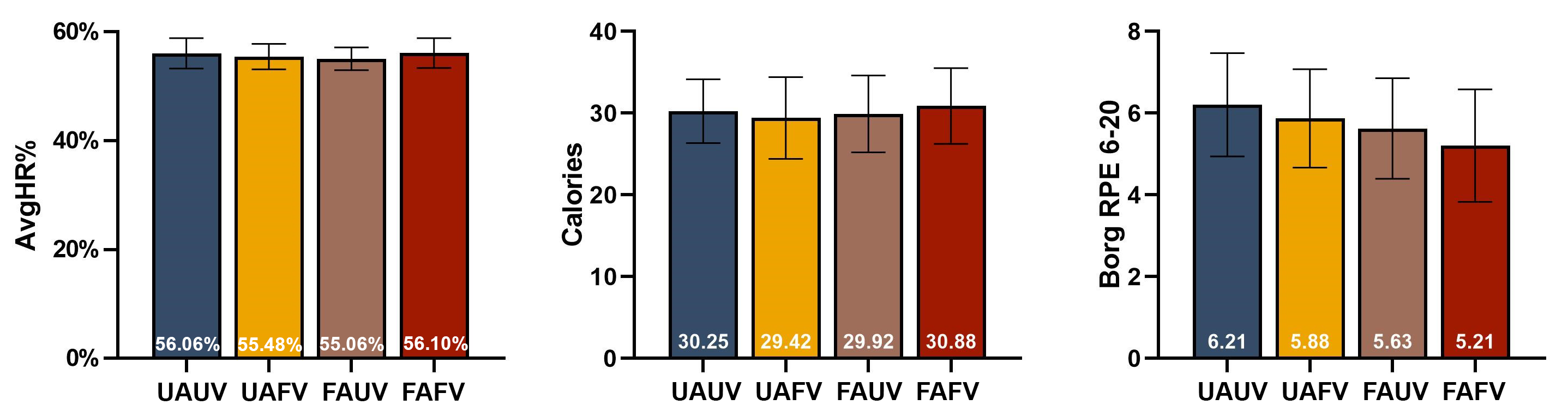}
    \caption{AvgHR\%, Calories burned, and Borg ratings for each condition. Error bars indicate 95\% confidence intervals.}
    \label{Exertion}
    
\end{figure*}

\subsection{Exertion}
Values for each condition for exertion measures can be found in Figure \ref{Exertion}. ANOVA tests revealed significant main effects of Audience Appearance ($F_{1,23}=5.374, p=.030, \eta_p^2=0.189$) and Audience Voice ($F_{1,23}=5.979, p=.015, \eta_p^2=0.231$) on perceived exertion (Borg RPE; after data were transformed via ART). No significant interaction effects were found ($F_{1,23}=0.702, p=.411$). Post-hoc pairwise comparisons indicated that participants experienced a higher level of exertion when (1) exposed to unfamiliar audience appearance ($M=6.02, SE=0.59$) compared to familiar audience appearance ($M=5.34, SE=0.59$)(2) confronted with unfamiliar audience voice ($M=5.98, SE=0.60$) than familiar audience voice ($M=5.44, SE=0.57$). We did not observe any significant main or interaction effects on avgHR\% and calories burned. 

Pearson's correlation revealed a significant correlation between participants' self-consciousness and their avgHR\% when exposed to unfamiliar audience appearance ($r=.428, p=.037$) and unfamiliar audience voice ($r=.459, p=.024$).
%Apart from heart rate, we did not find any significant correlations in other measures. These results suggest that individuals with higher self-consciousness levels may experience increased physiological responses, as reflected by elevated heart rates, in the presence of unfamiliar audience stimuli.

\subsection{User Rankings and Feedback}
Out of the 24 participants, 21 ranked Familiar Appearance with Familiar Voice (FAFV) as the best condition while the remaining 3 participants chose each of the other three conditions respectively. For the second place, 14 participants ranked Unfamiliar Appearance with Familiar Voice (UAFV) as their second choice, while 7 participants ranked Familiar Appearance with Unfamiliar Voice (FAUV) as their second choice. When asked about which factor (appearance or voice) had a more significant effect, 17 participants indicated that they were more easily influenced by audience voice, particularly when they were exposed to familiar audience voice.

When queried about additional details pertaining to their gaming experiences, 17 participants (P1-5, P11-P17, P19, P21-24) expressed a proclivity towards familiar elements, perceiving games featuring these elements as more \emph{``enjoyable/interesting"}. 7 participants (P3-P5, P8, P10, P16, P22) conveyed that the presence of familiar elements instilled a sense of \emph{``comfort/companionship"} during gameplay. Moreover, 4 participants (P9, P15, P19-P20) reported feeling \emph{``relaxed"} when confronted with familiar elements. Notably, a subset of 14 participants (P4-P6, P8-P10, P12-P13, P15, P18-P19, P21, P23-P24) opined that familiar elements fostered \emph{``focus"} and \emph{``enhanced performance,"} albeit 4 participants (P1, P3, P11, P22) perceived such elements as \emph{``distracting"} and impeding their gaming experience.

There was a mixed attitude towards positive and negative feedback, 7 participants (P1-P2, P5, P7, P9, P14, P17) perceived positive feedback as having a positive impact, whereas negative feedback was regarded as having a negative impact. In contrast, 6 participants (P3-P4, P13, P15-P16, P23) recognized both types of feedback being positive and identified negative feedback as a stimulus for \emph{``competitiveness"}. 6 participants (P6, P8, P10, P12, P18, P21) affirmed that both positive and negative feedback from familiar audience served as \emph{``motivation,"} whereas feedback from unfamiliar sources elicited no impact.

Regarding the potential of familiar audiences in future VR exergames to promote long-term engagement, 19 participants expressed a positive outlook. They conveyed a desire for NPC audiences that could be \emph{``customized"} and demonstrate heightened \emph{``intelligence."} However, the remaining 5 participants (P7, P13-P14, P17, P22) who responded negatively held the view that the sense of \emph{``comfort"} and \emph{``novelty"} elicited by familiar audiences was only ephemeral.

\section{Discussion}
\subsection{Impact of Appearance Familiarity on Player Performance and Experience}
To address \textbf{RQ1}, our research findings indicate that while appearance familiarity does not significantly impact player performance, it does demonstrate effects in enhancing player experience, promoting positive emotions, and reducing perceived exertion levels. 
% Participants reported feeling more engaged when interacting with appearancely familiar NPC audiences. Furthermore, appearance familiarity helped alleviate negative emotions associated with physical exertion, creating a more positive and comfortable atmosphere. 
This aligns with prior research findings \cite{macintyre1995effects}, which indicate that individuals tend to feel more relaxed and pleasant and have enhanced experiences when facing familiar audiences. These positive effects can be attributed to the familiarity and social connection established between the individuals and the audience, creating a more comfortable and engaging atmosphere \cite{burkitt2019expressivity}. 

However, while previous studies \cite{monteiro2023effects, barrett2023exploring} have shown the influence of appearance familiarity with the audience on performance outcomes in VR speech training tools, our findings did not reveal a significant effect in the context of exergames. One possible explanation for this discrepancy is the nature of the visually dominant gameplay, which requires players to maintain a visual focus on the tasks at hand. Several participants (P2, P4, P7, P16) mentioned that at the start of the game, they would purposefully gaze at the NPC's faces, but as the game progressed, they would focus on the gameplay itself with occasional glances towards them. It is worth noting that individuals have limited attentional capacity \cite{hubert2011stretching}, and they may not have enough available cognitive resources to allocate to the appearance aspects of NPC audiences. 

\subsection{Impact of Voice Familiarity on Player Performance and Experience}
Regarding \textbf{RQ2}, we found that voice familiarity played a crucial role in enhancing both gameplay performance and subjective experiences. Participants demonstrated notable improvements in their total score, fruit cut rate, and maximum combo count when exposed to familiar voices of NPC audiences. Similar to the appearance aspect, familiar voices also have a positive impact on player experience, emotions, and perceived level of exertion. 

One possible reason for the influence of voice familiarity on both performance and experience can be attributed to the characteristics of audio stimuli. Audios are typically provided for shorter durations than visuals, enabling immediate and continuous auditory cues that offer a constant stream of feedback throughout the interaction \cite{westerberg2015categorizing}. Moreover, audios are emitted omnidirectionally and constantly, as our ears cannot be closed, ensuring effective feedback delivery regardless of user orientation or visual focus \cite{serafin2018sonic}. Voice feedback does not occupy valuable visual space, allowing users to fully utilize their visual attention for other important aspects of the interaction \cite{palani2014evaluation, bosman2023effect}. While previous research \cite{monteiro2023effects, barrett2023exploring, raja2017anxiety} on audience familiarity in virtual environments has rarely involved voice familiarity, our findings indicate that the presence of familiar audience voice can have a meaningful impact on both the performance and overall engagement of users in VR exergames.

\subsection{Relationship between Self-Consciousness and the Impact of Audience Familiarity}
To answer \textbf{RQ3}, while there was a limited association between users' self-consciousness and audience familiarity in terms of performance, there was a significant correlation in terms of experience and average heart rates. Our results indicate that the presence of NPCs does not appear to impose pressure on participants with higher levels of self-consciousness, nor does it significantly impact their performance. However, the inclusion of familiar NPCs enhances their subjective experience and contributes to a more relaxed feeling. Interestingly, in the presence of unfamiliar audience stimuli, participants with higher self-consciousness levels may experience increased physiological responses, as reflected by elevated heart rates. On the other hand, they felt a greater sense of relatedness and competence when interacting with familiar audiences. The presence of familiar audiences seemed to provide them with a sense of comfort and social support. 

The results showed that participants with higher levels of self-consciousness demonstrated a greater sensitivity to the perception of audience familiarity, both in terms of appearance and voices. This finding aligns with previous research \cite{wang2004self, tice1985development} that suggests individuals with higher self-consciousness are more attentive to social cues and the perception of others' judgments. Previous research \cite{kappen2014engaged, haller2019hiit, yu2023cheer} has shown that the presence and feedback of an audience can enhance player performance and experience with limited exploration of players' self-consciousness. Our findings indicate that for individuals with self-conscious tendencies, the presence of familiar audiences may have a positive influence, resulting in a more relaxed and enjoyable gaming experience. 

\subsection{Design Guidelines}
\subsubsection{Incorporating Audience Familiarity into Exergames}
One of the key findings of our study is the positive impact of incorporating audience familiarity into gameplay. By integrating familiar NPC audiences, players not only improved performance but also enjoyed a heightened sense of game experience and positive emotions. Notably, our findings shed light on the influence of personality traits, particularly self-consciousness, on the effects of audience interactions. Understanding the role of personalities in audience interactions enables developers to tailor the game experience to individual differences.

To facilitate familiarity, we recommend the inclusion of customizable features (1. appearance---face, height, and body shape and 2. voice) for NPC audiences in VR exergames. By allowing players to import and incorporate their own network of acquaintances into the game, such as their friends and family members, NPC avatars can resemble familiar individuals from their personal lives. This recommendation acknowledges the significance of individual differences and empowers players to enhance their enjoyment and immersion in VR exergames.

\subsubsection{Prioritizing Voice Familiarity}
We recommend prioritizing integrating familiar voices in the design of NPC audiences in VR exergames. Voice familiarity played a pivotal role in influencing player performance and experience in VR exergames, while appearance familiarity only impacts experience. In addition, voice-only familiarity has a higher rating than appearance-only familiarity based on the ranking data and participants' interview feedback. Most importantly, the key advantage of emphasizing voice familiarity is the relative ease of capturing and incorporating familiar voices compared to facial features. Voice recordings can be obtained with greater convenience and accuracy, facilitating seamless integration into the game. This accessibility empowers games to easily leverage the power of familiar voice cues, enabling games with a more personalized and engaged gameplay experience for players.

\subsection{Limitations and Future Work}
Our study has provided valuable insights into the impact of audience familiarity on player performance and experience in VR exergames. However, there are opportunities for further exploration and improvement that can build upon our findings. One potential direction for future research is to expand the range of variables considered beyond facial appearance and voice. In particular, incorporating other elements, such as facial expressions and clothing, can offer a more comprehensive understanding of how different aspects of audience familiarity influence player experiences. 

Another area of improvement lies in the feedback provided by the NPC audience. While our study included a limited set of voice prompts and animations, there is room to enhance the immersive experience by incorporating a broader range of vocal expressions and animated responses. By allowing for customized vocal expressions and animations, future research can enable more dynamic and personalized interactions between players and the NPC audience. 
% This would not only enhance the sense of presence and engagement within the virtual environment but also provide users with a richer interactive experience.

The study participants mainly came from a young population. Exploring a wider range of age groups and demographic characteristics in the future will enhance the overall comprehensiveness of our findings. Furthermore, our study was conducted as a short-term experiment, limiting our understanding of the long-term effects of audience familiarity. Future research could address this limitation by conducting longitudinal studies that examine the sustained effects of audience familiarity over an extended period, such as a 6-week experiment in \cite{xu21UniStudents}. This would provide valuable insights into how familiarity with the virtual audience develops and evolves over time and how it impacts player engagement and performance in the long run. Additionally, while our results did not reveal significant issues with appearance and auditory mismatches---as evidenced by participants' ranking preferences---further investigation into factors like player acceptance and familiarity perception could be beneficial.

%In summary, while our study has shed light on the influence of audience familiarity in VR exergames, there are opportunities for future research to further enhance and expand upon our findings. By considering additional variables, improving interactive feedback, and exploring the long-term effects of audience familiarity, researchers can continue to advance our understanding of how virtual audiences can enhance player experiences in immersive gaming environments.

\section{Conclusion}
Our research illuminates the importance of audience familiarity in the context of VR exergames. We have demonstrated that both appearance and voice familiarity contributes to the player experience and perceived exertion, with voice familiarity exerting a more prominent influence on player performance. The positive impact of familiar NPC audiences on player experiences, particularly in terms of creating a relaxed and enjoyable atmosphere, suggests the potential for enhancing engagement and satisfaction in virtual reality environments. Furthermore, we discovered that individuals with higher self-consciousness are more sensitive to the influence of audience familiarity. This underscores the importance of considering individual differences and personal traits when designing virtual environments to ensure optimal player engagement and comfort. By emphasizing the integration of appearance and voice familiarity in NPC audience design, our study contributes to the advancement of immersive VR exergames that cater to individual preferences and promote positive player outcomes. Our findings offer valuable insights to developers, designers, and researchers in creating engaging VR exergames that harness the power of audience familiarity.

\acknowledgments{
The authors want to express gratitude to the participants who took part in the study, as well as to the reviewers, for their valuable insights and constructive suggestions that helped improve the paper.}

\bibliographystyle{abbrv-doi}

\bibliography{template}
\end{document}